\documentclass[pre,aps,showpacs]{revtex4}
\usepackage{epsfig}
\usepackage{bm}

\begin{document}

\title{Nonlocal interactions versus viscosity in turbulence}

\author{\small  A. Bershadskii}
\affiliation{\small {\it ICAR, P.O.\ Box 31155, Jerusalem 91000, Israel}}

\begin{abstract}
It is shown that nonlocal interactions determine energy spectrum
in isotropic turbulence at small Reynolds numbers. It is also shown that for
moderate Reynolds numbers the bottleneck effect is determined by the same
nonlocal interactions. Role of the large and small scales covariance at the 
nonlocal interactions and in energy balance has been investigated. A possible 
hydrodynamic mechanism of the nonlocal solution instability at large scales has 
been briefly discussed. A quantitative relationship between effective strain of the 
nonlocal interactions and viscosity has been found. All results are supported by 
comparison with the data of experiments and numerical simulations.
\end{abstract}

\pacs{47.27.-i, 47.27.Gs}

\maketitle

\section{Introduction}
Classic wind-tunnel experiments \cite{cc} showed that there is no scaling behavior
in a presumably  isotropic turbulence at {\it low} Reynolds numbers. Moreover, recent
high-resolution numerical simulations \cite{sys} performed for low-Reynolds-number
($R_{\lambda} \approx 10-60$) show no hint of scaling-like behavior of the velocity
increments even when ESS \cite{ben1} is applied. For {\it moderate} Reynolds numbers
numerical simulations show 'excess' power just before the dissipation (Kolmogorov's)
wavenumber $k_d$ (a hump in the compensated energy spectra), see for instance
\cite{gf}. This non-scaling effect (bottleneck effect \cite{f}) is usually
related with reducing efficiency of the energy cascade toward $k_d$ \cite{f},\cite{lm},\cite{vd}.
For the small Reynolds numbers applicability of
the energy cascade idea is problematic for entire range of scales. It is shown in recent paper
\cite{bersh} that nonlocal interactions become dominating in comparison with the local ones just
in the near-dissipation range of scales (cf. also \cite{dom}). In this range the viscous effects cannot be
neglected and scaling asymptote corresponding to the nonlocal regime cannot be observed \cite{bersh}
(in the inertial range the local interactions are presumably dominating ones and the nonlocal scaling
asymptote also cannot be observed). However, the nonlocal scaling asymptote can be used as a zeroth
term in a perturbation approach taking into account the viscosity effects. There are many ways
to develop such perturbation approach. For instance, in the paper \cite{bersh} a perturbation
approach giving logarithmic corrections to the scaling was developed, provided by significant role
of the kink instabilities of the vortex filaments at moderate and large values of Reynolds number
\cite{sb1}. This 'logarithmic'-perturbation approach is shown to be effective in a vicinity of the
crossover scale $r_c$, where exchange of stability between local and nonlocal regimes takes place \cite{bersh}.
This vicinity is rather wide at moderate values of Reynolds number, when an overlap between
these regimes is a strong phenomenon \cite{bersh}. For small Reynolds numbers, however, the kink
instabilities of the small vortex tubes is suppressed by the strong viscosity (see, for instance \cite{wood}
and references therein). Therefore, for the small Reynolds numbers another, adequate just for this case,
perturbation approach should be developed.
An approach of such type is suggested in present paper. 
Starting from this approach and using comparison with results of numerical 
simulations \cite{gfn} the laboratory experiments \cite{cc} it is shown that energy 
spectrum for small Reynolds numbers is determined by 
the nonlocal interactions even in isotropic turbulence. The same nonlocal interactions provide a 
hydrodynamic mechanism for the so-called bottleneck effect for moderate Reynolds numbers. 
It is also shown that large and small scales covariance at the nonlocal interactions plays 
a significant role in these phenomena. A quantitative relationship between effective strain of the 
nonlocal interactions and viscosity has been found using dynamical equations. 

\begin{figure} \vspace{-0.7cm}\centering
\epsfig{width=.7\textwidth,file=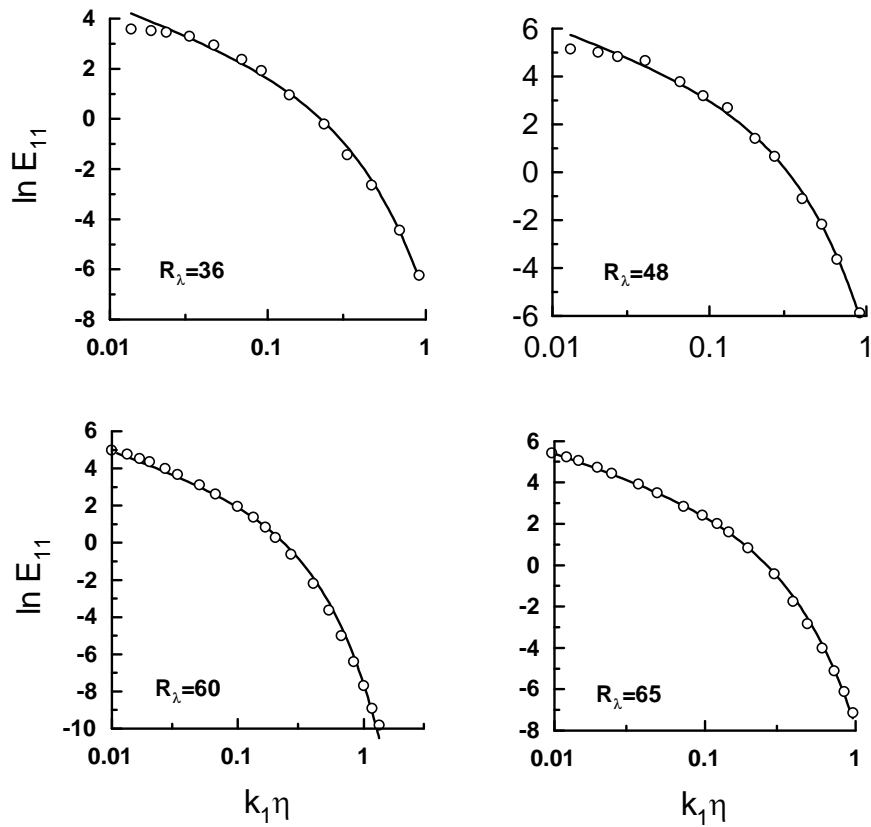} \vspace{-5cm}
\caption{One-dimensional spectra measured in nearly isotropic turbulence
downstream of a grid at small Reynolds numbers (the data are
reported in \cite{cc}). The solid curves are drawn in the figure to
indicate correspondence of the data to the equation (10)
(non-local regime). }
\end{figure}

\section{ Perturbations to scaling}

Let us following to the paper \cite{bersh} consider a dimensional function $E(k)$ of a dimensional
argument $k$. And let us construct a {\it dimensionless} function
of the same argument
$$
\alpha (k) = \frac{E^{-1}dE}{k^{-1}dk}.  \eqno{(1)}
$$
If for $k_d \gg k $ we have no relevant fixed scale (scaling
situation), then for these values of $k$ the function $\alpha (k)$
must be independent on $k$, i.e. $\alpha (k) \simeq const$ for $k_d \gg k $.
For turbulence $k$ could be a wavenumber and $k_d$ could be a dissipation wavenumber ($k_d =1/\eta$,
where $\eta = (\nu^3/\langle \varepsilon \rangle)^{1/4} $ is so-called viscous scale \cite{my}).
Solution of equation (1) with constant $\alpha$
can be readily found as
$$
E(k) \cong c k^{\alpha}    \eqno{(2)}
$$
where $c$ is a dimension constant. This is the well-known power
law corresponding to the scaling situations.

Let us now consider an analytic approach, which allows us to find corrections
of all orders to the approximate power law, related to the fixed scale $k_d$.
In the non-scaling situation let us denote
$$
f \equiv \ln (E/A), ~~~~~~~~ x \equiv \ln (k/k_d)  \eqno{(3)}
$$
where $A$ and $k_d$ are dimensional constants used for
normalization.

In these variables, equation (1) can be rewritten as
$$
\frac{df}{dx}=\alpha (x) \eqno{(4)}.
$$
In the non-scaling situation $x$ is a dimensionless variable,
hence the dimensionless function $\alpha (x)$ can be non-constant.
Since the 'pure' scaling corresponds to $k/k_d \ll 1$ we will use an
analytic expansion in power series
$$
\alpha (x) = \alpha_{0} + \alpha_1(k/k_d)+...
+ \frac{1}{n!}\alpha_n (k/k_d)^n+...  \eqno{(5)}
$$
where $\alpha_n$ are dimensionless constants. Choice of the small parameter
for analytic perturbation approach is determined by physical situation, which one
intends to consider. For instance, for {\it moderate} Reynolds numbers it would
be generally preferable to consider the $x^{-1}$ as a small parameter \cite{bersh}. In this
paper, however, we intend to consider energy spectra in isotropic turbulence at {\it small}
Reynolds numbers and corresponding phenomena in a relatively close vicinity of the
dissipation scale at moderate Reynolds numbers. The kink instabilities of the vortex
filaments, which are a significant factor near $r_c$ (see above and \cite{bersh})
are presumably not significant in situations with strong viscous effects \cite{wood}.
Therefore the choice of $x^{-1}$ does not seem to be relevant here. On the other hand,
parameter $k/k_d$ seems to be less related to the specific hydrodynamic structures
dominating processes in turbulence and more relevant to a sheer taking into account
corrections to the scaling providing by viscosity.

\begin{figure} \vspace{-0.4cm}\centering
\epsfig{width=.75\textwidth,file=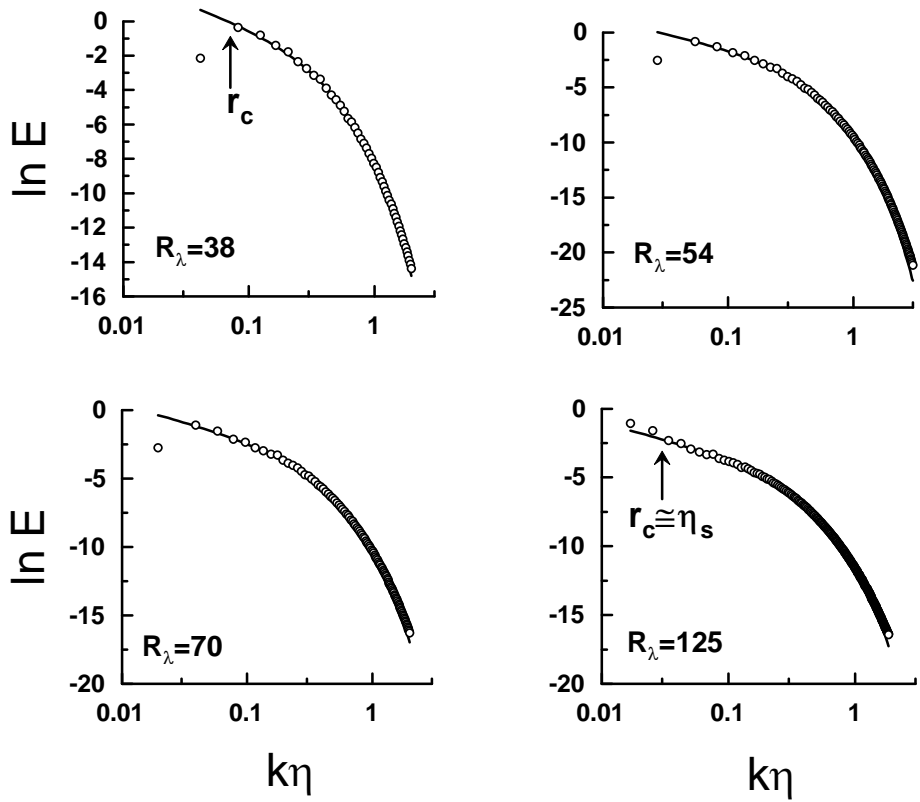} \vspace{-6.5cm}
\caption{Three-dimensional spectra from a DNS performed in \cite{gfn}
for different Reynolds numbers up to $R_{\lambda}=125$. The solid curves in this figure
corresponds to the best fit by equation (10) (nonlocal interactions). 
The value of $\alpha_1 =6.0 \pm 0.1$ is the same in all cases. }
\end{figure}

After substitution
of the analytic expansion (5) into Eq.\ (4) the zeroth order
approximation gives the power law (2) with $\alpha \cong
\alpha_0$. First order analytic approximation, when one takes only
the two first terms in the analytic expansion (5), gives
$$
E(k) \cong c k^{\alpha_0} e^{\alpha_1 k/k_d}.  \eqno{(6)}
$$
(cf , for instance, \cite{s1}-\cite{nelkin}). Corrections of the
higher orders can be readily found in this perturbation approach.

\section{ Nonlocal interactions}

Let us recall that in isotropic turbulence a complete separation of local and
non-local interactions is possible in principle. It
was shown by Kadomtsev \cite{kad} that this separation plays a
crucial role for the local Kolmogorov's cascade regime with scaling energy spectrum
$$
E(k) \simeq K~ \langle \varepsilon \rangle^{2/3} k^{-5/3}  \eqno{(7)}
$$
where $\langle \varepsilon \rangle$ is the average of the
energy dissipation rate, $\varepsilon$, $k = 1/r$ is the wave-number, and $K$ is the so-called Kolmogorov 
constant. This
separation should be effective for the both ends. That is, if there
exists a solution with the local scaling (7) as an asymptote, then there should also
exist a solution with the non-local scaling asymptote. Of course, the two solutions with
these asymptotes should be alternatively stable (unstable) in different regions
of scales. It is expected, that the local (Kolmogorov's) solution is stable (i.e.
statistically dominating) in inertial range (that means instability of the non-local
solution in this range of scales).

Roughly speaking, in non-local solution for small scales $r$ only
non-local interactions with large scales $L$ ($1 \gg r/L$) are
dynamically significant (the interaction among the small scales is negligible compared
with interaction via large scales) and the non-local interactions is
determined by large scale strain/shear. This means that one should
add to the energy flux $\langle \varepsilon \rangle$-parameter (which is a
governing parameter for the both solutions) an additional
parameter such as the strain $s$ for the non-local solution. As
far as we know it was noted for the first time by Nazarenko and
Laval \cite{nl} that dimensional considerations applied to
the non-local asymptote result in the power-law energy spectrum
$$
E(k) \simeq c~\frac{\langle \varepsilon \rangle}{s}~ k^{-1}  \eqno{(8)}
$$
both for two- and three-dimensional cases. Linear dependence of the
spectrum (8) on $\langle \varepsilon \rangle$ is determined by the
linear nature of equations corresponding to the non-local asymptote
that together with the dimensional considerations results in (8)
\cite{nl}. Interesting numerical simulations were performed in
\cite{ldn}. In these simulations local and non-local interactions
have been alternatively removed. For the first case a tendency
toward a spectrum flatter than '-5/3' is observed near and beyond
the separating scale  (beyond which local interactions are ignored),
that supports Eq. (8).

\begin{figure} \vspace{-1cm}\centering
\epsfig{width=.55\textwidth,file=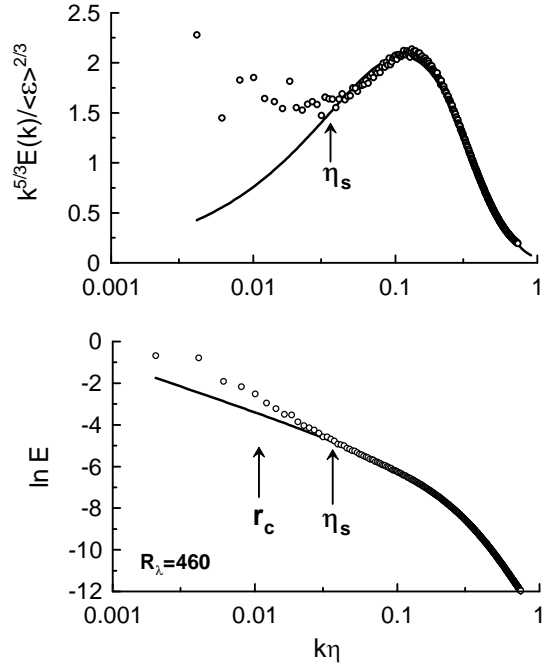} \vspace{-3cm}
\caption{As in Fig. 2 but for $R_{\lambda}=460$. The upper part of the figure shows
the same energy spectrum as the lower one but in the compensate (according to the Kolmogorov's scaling) form.
The solid curves are the best fit corresponding to the nonlocal approximation (10) with the same
$\alpha_1 =6.0 \pm 0.1$ as in Figs 2.}
\end{figure}

Following to the perturbation approach suggested above both
local and non-local regimes can be corrected. The first order correction is
$$
E(k) \simeq K \langle \varepsilon \rangle^{2/3}
k^{-5/3} e^{- \beta (k/k_d)}  \eqno{(9)}
$$
and
$$
E(k) \simeq c \frac{\langle \varepsilon \rangle}{s} k^{-1} e^{- \alpha_1 (k/k_d)}  \eqno{(10)}
$$
for the local and non-local regimes respectively ($K$, $c$, $\alpha_1$ and
$\beta$ are dimensionless constants).

Figure 1 shows one-dimensional spectra measured in nearly isotropic turbulence
downstream of a grid at small Reynolds numbers (the data are
reported in \cite{cc}). The solid curves are drawn in the figure to
indicate correspondence of the data to the equation (10)
(non-local regime). The experimental data, however, is not controlled enough to
study fine properties of the isotropic turbulence. Therefore below we will mainly use
the data obtained in numerical simulations.

It is shown in \cite{bersh} that there is an 'exchange of stability'
phenomenon at certain $k_c=1/r_c$. That is, for $k < k_c$ the
Kolmogorov's regime is stable and the non-local regime is unstable,
whereas for $k > k_c$ the Kolmogorov's regime is unstable and the
non-local regime is stable. For this scenario, at $k=k_c$ the
Kolmogorov's regime is still asymptotically scale-invariant (i.e.
Eq.\ (7) gives an adequate approximation for this regime), while for
the non-local regime the first order correction is substantial (i.e.
Eq.\ (10) should be used at $k > k_c$ for the non-local regime). In
this scenario the Kolmogorov's regime plays significant role in the
viscous stabilization of the non-local regime for $k > k_c$ (see
\cite{ldn}), but for these $k$ the non-local regime becomes
statistically dominating instead of the Kolmogorov's one (it is well
known that only for energy spectra steeper than $k^{-3 }$ it can be
rigorously proved that the dominant interactions are nonlocal (cf
Eq. (10) that provides such steepness).

It should be noted that in \cite{yz} a second scaling solution $E(k)
\propto P k^{-1}$ was obtained in addition to the Kolmogorov scaling
using the Clebsch formulation of hydrodynamics. The prefactor $P=
(\langle \varepsilon \rangle \nu)^{1/2}$ denotes a certain flux in
the wavenumber space \cite{yz}. It will be shown below that the 
prefactor in Eqs. (8),(10) can be transformed into the one formally equal to
$P$ (see Eq. (15) and Fig. 4). This could mean that the studied nonlocal regime
(more precisely: its scaling asymptote) is closely related to the 
Yakhot-Zakharov scaling solution introduced in the
Ref. \cite{yz}. Such identification (if valid) allows us understand
instability of the nonlocal regime for large scales (small $k$). Indeed, 
it is shown in \cite{n} that reconnection process breaks conservation
of the integral determining the flux $P$ for the large scales (small
$k$) and , therefore, there is no possibility for realization of the
scaling asymptote $k^{-1}$ of this solution. Following to Newell
\cite{n}\cite{n2} the nonconservation of the integral in large
scales generally follows from nonlocality of the viscosity term in
the equations formulated for the Clebsch variables. Since for the
large scales the nonlocal regime should be represented by its
asymptote $E \propto k^{-1}$ (as an intermediate asymptote \cite{bar} 
the nonconservation of the integral
makes this regime unstable for the large scales (small $k$).

Figure  2 shows three-dimensional energy spectra calculated using
data from a high-resolution direct numerical simulation of
homogeneous steady three-dimensional turbulence \cite{gfn} for
different Reynold numbers up to $R_{\lambda}=125$. The solid curves
in this figure corresponds to the best fit by equation (10)
(nonlocal interactions). The exponent $\alpha_1 \simeq 6.0 \pm 0.1$
for all considered values of $R_{\lambda}$. The arrows indicate the
$r_c=1/k_c$ scales calculated using corresponding $D_{LLL} (r)$ (see
\cite{bersh}).

In figure 3 we show the data obtained in the same DNS as those shown in Fig. 2 but for
$R_{\lambda}=460$. In the upper part of this figure we show the energy spectrum in the compensate
(according to Kolmogorov's scaling Eq. (7)) form. One can clear see the hump corresponding to the
bottleneck effect. The solid curves are the best fit to the the same nonlocal spectrum (10) with
the same (universal) value of $\alpha_1 \simeq 6.0 \pm 0.1$ (see also below). 
One can see that the bottleneck effect
is determined by just the same nonlocal interactions as the above considered energy spectra for small
Reynolds numbers.

\section{ Strain and viscosity}
\begin{figure} \vspace{-1cm}\centering
\epsfig{width=.55\textwidth,file=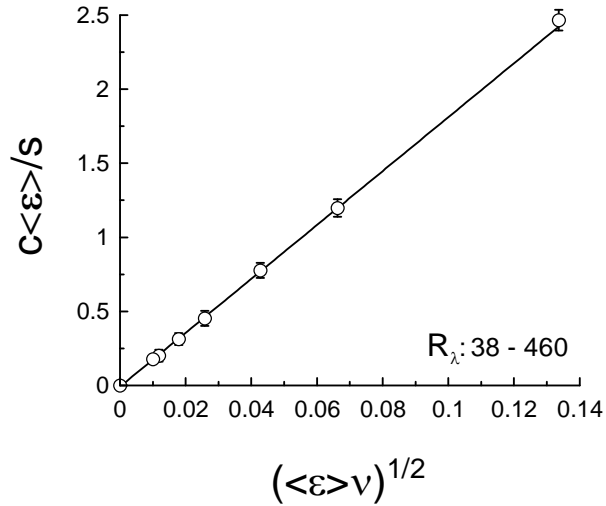} \vspace{-5.5cm}
\caption{The prefactor in the nonlocal approximation to the energy spectra (10): 
$c \langle \varepsilon \rangle/s$ - circles, versus $(\langle
\varepsilon \rangle \nu)^{1/2}$. The straight line with the slope equals to 
18 indicates agreement with Eq. (15).}
\end{figure}

It is shown in 
Ref. \cite{bersh} that for sufficiently large Reynolds numbers, providing
a visible inertial interval, there is an overlapping between the two regimes:
non-local and local (Kolmogorov). This overlapping is based on the very nature
of the stability exchange between the two {\it statistical} regimes.
However, in analogy with the viscous scale $\eta =  (\nu^3/\langle \varepsilon
\rangle)^{1/4} $ \cite{my} there should be a scale $\eta_s$ such that for scales $r < \eta_s$
contribution of the local interactions will be drastically decreased in comparison with the nonlocal
ones. In the analogy with $\eta$ one can calculate $\eta_s$ 
using the dimensional considerations as
$$
\eta_s = \left(\frac{\langle \varepsilon \rangle}{s^3}\right)^{1/2}  \eqno{(11)}
$$
(cf for shear flows \cite{tas}).

 The dynamical equations provide us with a relationship \cite{my}
$$
\langle \varepsilon \rangle = 2\nu \int^{k_d}_{0} k^2 E(k) dk \eqno{(12)}
$$
For $E(k)$ given by Eq. (10) the dissipation function $k^2E(k)$ has its maximum at $k=k_d/6$. 
If $k_d \gg k_s=1/\eta_s$ (see below) the maximum of the dissipation function $k^2E(k)$ 
is located just between $k_s$ and $k_d$. Therefore, we can estimate (12) as
$$
\langle \varepsilon \rangle \simeq 2\nu \int^{k_d}_{k_s} k^2 E(k) dk \eqno{(13)}
$$
Substituting (10) (with $\alpha_1 =6$) into (13) we obtain relationship
$$
s \simeq \frac{c}{18} \langle \varepsilon \rangle^{1/2} \nu^{-1/2}  \eqno{(14)}
$$

Using the relationship (14) one can 
estimate the prefactor in the
nonlocal approximation to the energy spectra (10) as
$$
c \frac{\langle \varepsilon \rangle}{s} \simeq 18 (\langle
\varepsilon \rangle \nu)^{1/2}  \eqno{(15)}
$$
Now using the data of the DNS \cite{gfn} (cf Figs. 2,3) let us calculate 
the prefactor $c \langle \varepsilon \rangle/s$. Results of these calculations
are shown in figure 4 as circles ($Re_{\lambda}= 38, 54, 70, 125,$ $284, 380, 460$). 
This figure shows the prefactor $c ~\langle \varepsilon \rangle/s$ against 
$(\langle \varepsilon \rangle\nu)^{1/2}$. The straight line with the slope equals to 18 
indicates agreement with Eq. (15). 
 
It should be noted that in the DNS \cite{gfn}
$\langle \varepsilon \rangle \simeq const$ for $R_{\lambda}\geq 70$,
in agreement with the well known Kolmogov's hypothesis \cite{my}. Therefore, for $R_{\lambda} \geq
70$ (when $\langle \varepsilon \rangle \simeq const$ \cite{gfn}) it follows
from Eq. (14) that $s \propto \nu^{-1/2}$. This relationship provides us 
also with dependence of the strain $s$ on $R_{\lambda}$. \\

Due to the nonlinear character of the Navier-Stokes equations the so-called
triadic type of interactions is dominating mechanism of the dynamical interactions
in turbulence (see, for instance, \cite{dom}). A triad corresponding to the
nonlocal interactions involves two short-wave-number modes and one long-wave-number mode
(see a sketch in figure 5). Accordingly, at the nonlocal interactions {\it two} characteristic
space scales are actively involved: large-scale characteristic scale $\eta_s$ (11) and
small-scale (viscous or Kolmogorov) characteristic space scale $\eta =(\nu^3/\langle \varepsilon
\rangle)^{1/4} $. It is naturally that the large scales should be normalized by $\eta_s$ while
the small scales should be normalized by $\eta$. This could cause an obvious problem at the
nonlocal interactions. However,
the large and small scales covariance relationship
at the nonlocal interactions follows directly from the relationship (14)
$$
\frac{\eta_s}{\eta} = \left( \frac{\langle \varepsilon \rangle}{\nu s^2 }\right)^{3/4} 
\simeq \left( \frac{18}{c} \right)^{3/2} \simeq const   \eqno{(16)}
$$
(i.e. the relation $\eta_s/\eta$ is independent on $R_{\lambda}$). And vice versa, the
large and small scales covariance (15) results in the relationship of the type (14). The covariance
at the nonlocal interactions supports right balance between the energy flux to the small scales
and their dissipative capacity (cf, for instance, \cite{dg},\cite{p}).

\begin{figure} \vspace{-0.5cm}\centering
\epsfig{width=.45\textwidth,file=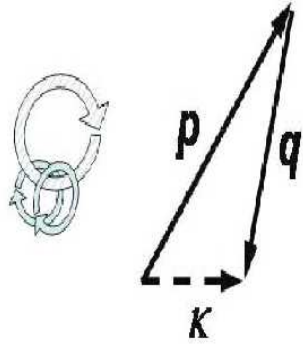} \vspace{-5cm}
\caption{Sketch of the nonlocal triadic interaction in the physical and in the Fourier space}
\end{figure}

Actually, the scale $\eta_s=1/k_s$ should provide an edge of applicability of
the approximation (10) to the real energy spectra. If one
compares the $\eta_s$ determined by this way from the Figs. 2,3,6 one can see that indeed
$\eta_s/\eta \simeq const$ in agreement with Eq. (16) (we have also checked this with the data
for $R_{\lambda} = 284, 380$ \cite{gfn} and the data reported in \cite{kaneda}). 
Moreover, taking into account a continuity condition 
of the energy spectrum at the point $k=k_s$ (and using Eqs. (7) and (10)):
$$
K~ \langle \varepsilon \rangle^{2/3} k_s^{-5/3} \simeq c \frac{\langle \varepsilon \rangle}{s} 
k_s^{-1} e^{- \alpha_1 (k_s/k_d)}  \eqno{(17)}
$$
we obtain equation
$$
K \simeq c~\exp[-(2c)^{3/2}/\alpha_1^2]  \eqno{(18)}
$$
Substituting the Kolmogorov constant $K \simeq 1.6$ (see \cite{gfn},\cite{s2}) and $\alpha_1 \simeq 6$ 
into this equation we obtain $c \simeq 2$. Then, substituting this value of $c$ into Eq. (16) we obtain 
$k_s/k_d \simeq 0.037$. The last value is in agreement with the available data (see Figs. 2,3,6 and \cite{kaneda}). 

Together with the universality of $\alpha_1 \simeq 6.0$ 
(which also is a consequence of the
scale covariance) Eq. (16) determines the well
known from DNSs \cite{gfn},\cite{kaneda} universality (independence on $R_{\lambda}$) of the
position of the 'hump' in the axes $k\eta$ for the bottleneck effect.
\begin{figure} \vspace{-0.4cm}\centering
\epsfig{width=0.5 \textwidth,file=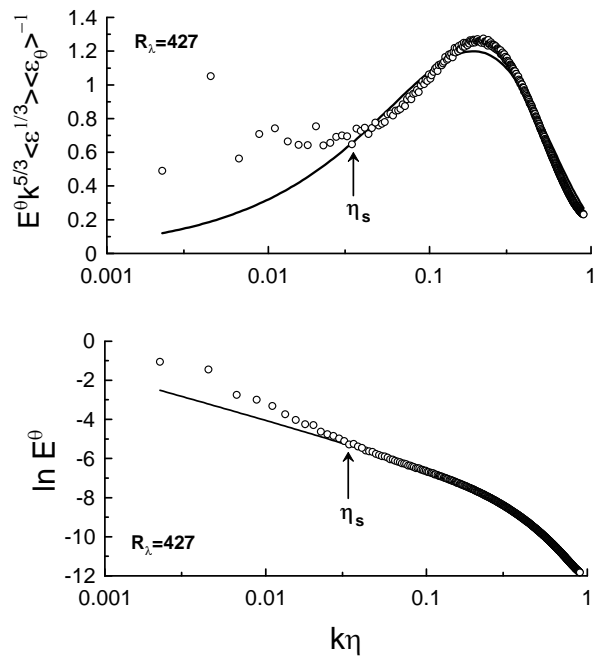} \vspace{-2.5cm}
\caption{ The passive scalar DNS data (circles) for homogeneous isotropic turbulence
described in \cite{wg}, P\'eclet number $P_{\lambda}=427$ and the Schmidt number is unity
(i.e $P_{\lambda}=R_{\lambda}$). The solid curves in this figure
correspond to the best fit by equation (20) (nonlocal interactions).}
\end{figure}

\section{Passive scalar}

For passive scalar $\theta$ in the isotropic turbulence the equations (9),(10) should be replaced
\cite{bersh} by the equations
$$
E^{\theta} (k) \propto \langle \varepsilon \rangle^{-1/3} \langle \varepsilon_{\theta} \rangle
~k^{-5/3} e^{- \gamma (k/k_d)}  \eqno{(19)}
$$
and
$$
E^{\theta} (k) \propto \frac{\langle \varepsilon_{\theta} \rangle}{s}~k^{-1} e^{- \delta (k/k_d)}  \eqno{(20)}
$$
for the local and non-local regimes respectively ($\gamma$ and $\delta$ are dimensionless
constants, and $\langle \varepsilon_{\theta} \rangle $ is the average value of dissipation
rate of scalar variance).

Figure 6 shows three-dimensional passive scalar spectrum from a DNS performed in \cite{wg}
for P\'eclet number $P_{\lambda}=427$ (the Schmidt number is unity,
i.e $P_{\lambda}=R_{\lambda}$). The solid curves in this figure
correspond to the best fit by equation (20) (nonlocal interactions).
In the upper part of this figure we show the spectrum in the compensate
(according to the Corrsin-Obukhov scaling \cite{wg}) form. One can clear see the hump
corresponding to the bottleneck effect. The arrows show position of the $\eta_s$ scale.
Comparing with Figs. 2,3 one can see that for the passive scalar $\eta_s/\eta$ takes the
same universal value as for the velocity field (the large and small scales covariance (15)).

\acknowledgments

I thank K.R. Sreenivasan for inspiring cooperation. I also thank T. Nakano, D. Fukayama and T. Gotoh for
sharing their data and discussions.


\begin{thebibliography}{99}
\bibitem{cc} G. Comte-Bellot and S. Corrsin, J. Fluid Mech., {\bf
48}, 273 (1971).
\bibitem{sys} J. Schumacher, K. R. Sreenivasan, and V. Yakhot, New Journal
of Physics {\bf 9}, 89 (2007).
\bibitem{ben1} R. Benzi, S. Ciliberto, R. Tripiccione, C. Baudet, F. Massaioloi, and 
S. Succi, Phys. Rev. E., {\bf 48}, R29 (1993).
\bibitem{gf} T. Gotoh and D. Fukayama, Phys. Rev. Lett. {\bf 86}, 3775
(2001).
\bibitem{f} G. Falkovich, Phys. Fluids {\bf 6}, 1411 (1994).
\bibitem{lm} D. Lohse and A. Muller-Groeling, Phys. Rev. Lett., {\bf 74},
1747 (1995); Phys. Rev. E {\bf 54}, 395 (1996).
\bibitem{vd} M.K. Verma and D. Donzis, J. Phys. A, {\bf 40}, 4401 (2007).
\bibitem{bersh} A. Bershadskii, J. Stat. Phys., online first: DOI 10.1007/s10955-007-9322-0 
(see also arXiv:nlin.CD/0603070).
\bibitem{dom} J.A. Domaradzki, Phys. Fluids A, {\bf 4}, 2037 (1992).
\bibitem{sb1} K.R. Sreenivasan and A. Bershadskii, J.\ Fluid.\ Mech.\, {\bf 554}, 477 (2006).
\bibitem{wood} P.R. Woodward, D.H. Porter, B. K. Edgar, S. E. Anderson,
and G. Basset, Comput. Appl. Math. {\bf 14}, 97 (1995).
\bibitem{gfn} T. Gotoh, D. Fukayama and T. Nakano, Phys.\ Fluids,
{\bf 14}, 1065 (2002).
\bibitem{s1} K.R. Sreenivasan, J. Fluid Mech., {\bf 151}, 81 (1985).
\bibitem{fms} C. Foias, O. Manley, and L. Sirovich, Phys. Fluids, A {\bf 2}, 464 (1990).
\bibitem{sj} Z-S. She and E. Jackson, Phys. Fluids A, {\bf 5} 1526
(1993).
\bibitem{nelkin} M. Nelkin, Adv.\ Phys. {\bf 43}, 143 (1994).
\bibitem{kad} B.B. Kadomtsev, Plasma Turbulence (Academic Press, New York, 1965).
\bibitem{nl} S. Nazarenko and J.-P. Laval, J. Fluid Mech., {\bf 408}, 301 (2000).
\bibitem{ldn} J-P. Laval, B. Dubrulle and S. Nazarenko, Phys.\ Fluids, {\bf 13}: 1995 (2001).
\bibitem{yz} V. Yakhot and V. Zakharov, Physica D, {\bf 64} 379 (1993).
\bibitem{n} S.V. Nazarenko, Physica D, {\bf 102} 343 (1997).
\bibitem{n2} A.C. Newell, private communication.
\bibitem{bar} G.I. Barrenblatt, Scaling, self-similarity, and intermediate asymptotics 
(Plenum Press, New York/London, 218 p., 1979).
\bibitem{my} A.S. Monin and A.M. Yaglom, Statistical Fluid Mechanics, Vol.\ 2,
(MIT Press, Cambridge 1975).
\bibitem{tas} F. Toschi, G. Amati, S. Succi, R. Benzi, and R. Piva, Phys. Rev. Lett., {\bf 82}, 
5044 (1999).     
\bibitem{dg} B. Dubrulle and J. Graner, J. Phys. II, {\bf 6}, 797 (1996).
\bibitem{p} A. Pocheau, Europhys. Lett, {\bf 35}, 183 (1996).
\bibitem{s2} K. R. Sreenivasan, Phys. Fluids {\bf 7}, 2778 (1995).
\bibitem{kaneda} T. Ishihara, Y. Kaneda, M. Yokokawa, K. Itakura and A. Uno,
J. Phys. Soc. of Japan, {\bf 74}, 1464 (2005).
\bibitem{wg} T. Watanabe and T. Gotoh, New J. Phys. {\bf 6}: Art. No. 40. (2004)
\end{thebibliography}
\end{document}